\documentclass[10pt,twoside,twocolumn,english,aps,manuscript,preprint,showpacs]{revtex4}
\usepackage[T1]{fontenc}
\usepackage[latin9]{inputenc}
\usepackage{units}
\usepackage{amstext}
\usepackage{graphicx}

\makeatletter
\@ifundefined{textcolor}{}
{%
 \definecolor{BLACK}{gray}{0}
 \definecolor{WHITE}{gray}{1}
 \definecolor{RED}{rgb}{1,0,0}
 \definecolor{GREEN}{rgb}{0,1,0}
 \definecolor{BLUE}{rgb}{0,0,1}
 \definecolor{CYAN}{cmyk}{1,0,0,0}
 \definecolor{MAGENTA}{cmyk}{0,1,0,0}
 \definecolor{YELLOW}{cmyk}{0,0,1,0}
 }

\makeatother

\makeatother

\usepackage{babel}

\begin{document}

\title{$^{119}$Sn NMR probe of magnetic fluctuations in SnO$_{2}$ nanoparticles}

\author{Tusharkanti Dey}

\email[Email: ]{tusdey@gmail.com}

\affiliation{Department of Physics, Indian Institute of Technology Bombay, Powai,
Mumbai 400076, India}

\author{P. Khuntia}

\affiliation{Department of Physics, Indian Institute of Technology Bombay, Powai,
Mumbai 400076, India}

\author{A.V. Mahajan}

\affiliation{Department of Physics, Indian Institute of Technology Bombay, Powai,
Mumbai 400076, India}

\author{Nitesh Kumar}

\affiliation{Jawaharlal Nehru Centre for Advanced Scientific Research (JNCASR),
Jakkur P. O., Bangalore 560064, India}

\author{A. Sundaresan}

\affiliation{Jawaharlal Nehru Centre for Advanced Scientific Research (JNCASR),
Jakkur P. O., Bangalore 560064, India}
\begin{abstract}
$^{119}$Sn nuclear magnetic resonance (NMR) spectra and spin-lattice
relaxation rate ($1/T_{1}$) in SnO$_{2}$ nanoparticles were measured
as a function of temperature and compared with those of SnO$_{2}$
bulk sample. A $15\%$ loss of $^{119}$Sn NMR signal intensity for
the nano sample compared to the bulk sample was observed. This is
indicative of ferromagnetism from a small fraction of the sample.
Another major finding is that the recovery of the $^{119}$Sn longitudinal
nuclear magnetization in the nano sample follows a stretched exponential
behavior, as opposed to that in bulk which is exponential. Further,
the $^{119}$Sn $1/T_{1}$ at room temperature is found to be much
higher for the nano sample than for its bulk counterpart. These results
indicate the presence of magnetic fluctuations in SnO$_{2}$ nanoparticles
in contrast to the bulk (non-nano) which is diamagnetic. These local
moments could arise from surface defects in the nanoparticles. 
\end{abstract}


\maketitle

\textbf{Introduction.} - The discovery of ferromagnetism even for
dilute doping in semiconductors and insulators has been one of the
biggest surprises for researchers. In 1998, H. Ohno \cite{Ohno-science-281-1998}
reported the observation of ferromagnetism in Mn doped GaAs with a
Curie temperature as high as 110K. After two years, T. Dietl \textit{et
al.} \cite{Dietl-science-287-2000} first predicted the possible appearance
of ferromagnetism above room temperature in Mn doped ZnO and GaN.
After that many researchers put great effort to find a room temperature,
intrinsic, dilute magnetic semiconductor (DMS). Thin films and bulk
samples of transition metal doped metal oxides like TiO$_{2}$, ZnO,
In$_{2}$O$_{3}$, SnO$_{2}$, and CeO$_{2}$ were found to show ferromagnetism
\cite{Sharma-Nat. Mater-2-2003,Hong-APL-87-2005,Tiwari-APL-88-2006,Ogale-PRL-91-2003,Matsumoto-science-291-2001}
at room temperature. More interestingly, thin films of a band insulator
HfO$_{2}$, which is diamagnetic in bulk form, was found to be ferromagnetic
without any doping, with Curie temperature greater than 500K \cite{Venkatesan-nature-430-2004}.
Thereafter, many \textit{undoped,} wide band gap, semiconductor metal
oxides such as TiO$_{2}$, ZnO, SnO$_{2}$, In$_{2}$O$_{3}$, Al$_{2}$O$_{3}$,
CeO$_{2}$, etc. were found to show room temperature ferromagnetism
in nanoparticulate or thin film form \cite{Sundaresan-PRB-74-2006,Banerjee-APL-91-2007,Hong-PRB-73-2006}.
Sundaresan \textit{et al.} \cite{Sundaresan-PRB-74-2006} suggested
that all metal oxides in nanoparticulate form would exhibit room temperature
ferromagnetism. All these discoveries of ferromagnetism in semiconducting
or insulating oxides without any unpaired `$d$' or `$f$' electron
have questioned the basic understanding of the origin of magnetism
\cite{Venkatesan-nature-430-2004}.

Sundaresan \textit{et al.} \cite{Sundaresan-PRB-74-2006} argued that
oxygen vacancies at the surfaces are responsible for ferromagnetism
in CeO$_{2}$, Al$_{2}$O$_{3}$, ZnO, In$_{2}$O$_{3}$, and SnO$_{2}$
nanoparticles. From \textit{ab initio} density functional calculations,
Ganguli \textit{et al.} \cite{Ganguli-JAP-108-2010} argued that defects
in ZnO clusters even in the absence of transition metal doping may
drive the cluster magnetic. Using density functional theory, Rahman
\textit{et al.} \cite{Rahman-PRB-78-2008} observed that in SnO$_{2}$
nanoparticles oxygen vacancy is nonmagnetic but a tin vacancy is ferromagnetic
with a large magnetic moment and coupled strongly with other Sn defects.
In another study by first principles calculation, Wang \textit{et
al.} \cite{Wang-PSS-247-2010} have concluded that surface defects
are responsible for ferromagnetism in undoped SnO$_{2}$ nanoparticles.
All these studies infer that defects at the surface play a key role
in ferromagnetism in SnO$_{2}$ nanoparticles.

Nuclear magnetic resonance (NMR) is widely used for detecting magnetic
fluctuations, crystal environment etc. at a local level in bulk materials.
NMR is also used to study nano materials \cite{Tomaselli-JCP-110-1999,Sabarinathan-JNN-8-2008,Tang-PRB-77-2008,Monredon-JMC-12-2002}.
Tang \textit{et al.} \cite{Tang-PRB-77-2008} have measured $^{67}$Zn
NMR in ZnS nanoparticles. They reported no detectable NMR signal due
to quantum size effects when the average particle size was below $4$\,nm
and signal was observed when average particle size was more than $8$\,nm.
Monredon\textit{ et al.} \cite{Monredon-JMC-12-2002} performed $^{119}$Sn
MAS NMR on nanoparticles of SnO$_{2}$ of different sizes to show
that the spectral shape change with the average size of the nanoparticles.
Sabarinathan \textit{et al.} \cite{Sabarinathan-JNN-8-2008} also
performed $^{119}$Sn MAS NMR on SnO$_{2}$ nanoparticles to measure
spectra as well as spin-spin and spin-lattice relaxations for two
samples of different particle size. Their study shows that smaller
the size of the particles, faster is the relaxation. They concluded
that inhomogenious broadening of the spectra is due to the surface
defects of the nanocrystalline samples. However, the spin-lattice
relaxation analysis by them is not reliable due to the lack of optimal
measurement parameters \cite{footnote}. According to our knowledge,
no detailed NMR measurement on ferromagnetic metal oxide nanoparticles
is reported so far.

With the motivation of verifying the presence of local moments in
nanoparticles via a local probe, we have chosen SnO$_{2}$ as a representative
candidate for detailed NMR measurements. We have measured $^{119}$Sn
NMR spectra and spin-lattice relaxation time at different temperatures
for the SnO$_{2}$ nano sample and compared them with those in the
SnO$_{2}$ bulk sample. We observed a $15\%$ loss of the $^{119}$Sn
NMR spectral intensity in the nano sample when compared with the bulk.
This is likely due to static, local magnetic moments on a fraction
of Sn atoms. This could very well be due to (possibly defect induced)
ferromagnetic surfaces of the smaller nanoparticles. The large local
field at these $^{119}$Sn nuclei will shift the resonance out of
the window of our observation. Non single-exponential recoveries obtained
for the nano sample in spin-lattice relaxation measurements are probably
due to the distribution of relaxation rates from different paramagnetic
layers in the nanoparticles. The faster relaxation rate for the nano
sample compared to the bulk sample indicates the presence of magnetic
fluctuations in the nano sample.

\textbf{Experimental details.} - SnO$_{2}$ nanoparticles were prepared
at JNCASR by a modified synthetic procedure discussed by Gnanam \textit{et
al.} \cite{Gnanam-Jsol-Gel-53}. In a typical synthesis, $2.5$\,g
of SnCl$_{2}$ was dissolved in ethanol to get a clear solution. $6\,$mL
ammonia ($30\%$) was added slowly to the above solution with constant
stirring to get a white precipitate. The precipitate was centrifuged
and washed with distilled water several times till the pH became $7$
followed by the last washing with ethanol. The white product was dried
at $80^{0}$\,C for $5$\,h. After drying, the product became yellow.
The yellow powder was heated at $450^{0}$\,C in oxygen atmosphere
for two hours to get crystalline nanoparticles of SnO$_{2}$. The
bulk sample was prepared by pelletizing the powder and sintering the
cylindrical pellet at $1100^{0}$\,C for $24$\,h.

X-ray diffraction (XRD) patterns were collected with a Rigaku-$99$
diffractometer using Cu $K_{\alpha}$ radiation ($\lambda=1.5406\textrm{\AA}$).
Transmission electron microscopy (TEM) images were recorded with a
JEOL JEM $3010$ instrument (Japan) operated with an accelerating
voltage of $300$\,kV. Magnetic measurements were carried out using
a Quatnum Design Physical Property Measurement System (PPMS).

After these basic characterizations, we have carried out $^{119}$Sn
NMR measurements for both bulk and nano samples of SnO$_{2}$ at IIT
Bombay. Sn has three NMR active isotopes ($^{115}$Sn, $^{117}$Sn,
and $^{119}$Sn). Our measurements are on the $^{119}$Sn nucleus
($I=1/2$) because it has the highest $\nicefrac{\gamma}{2\pi}$ value
($15.869$\,MHz/T) and the highest natural abundance ($8.58\%$)
among the three. For NMR measurements we have used a Tecmag pulse
spectrometer and a fixed magnetic field $93.954$\,kOe obtained inside
a room temperature bore Varian superconducting magnet. Variable temperature
measurements were carried out with the help of an Oxford cryostat
and accessories. Spin lattice relaxation was measured using a standard
saturation recovery method following a $\nicefrac{\pi}{2}-t-(\nicefrac{\pi}{2}-\pi)$
pulse sequence where the typical $\nicefrac{\pi}{2}$ pulse width
was $6\mu s$.

\begin{figure}
\centering{}\includegraphics[scale=0.3]{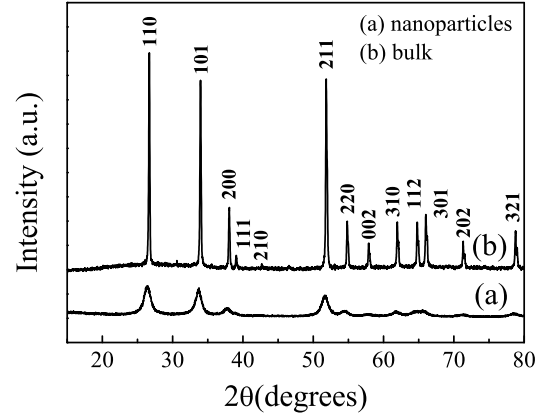}\caption{\label{fig:XRD} XRD patterns of (a) nanoparticles and (b) bulk SnO$_{2}$
sample.}

\end{figure}

\textbf{Results and discussions.} - Figure \ref{fig:XRD} shows XRD
patterns of nanoparticles as well as bulk SnO$_{2}$. The nanoparticles
as well as bulk powder crystallize in a tetragonal rutile structure.
Using Scherrer's formula for the XRD peak broadening we obtain the
particle size to be $\sim13$\,nm.

\begin{figure}
\centering{}\includegraphics[scale=0.5]{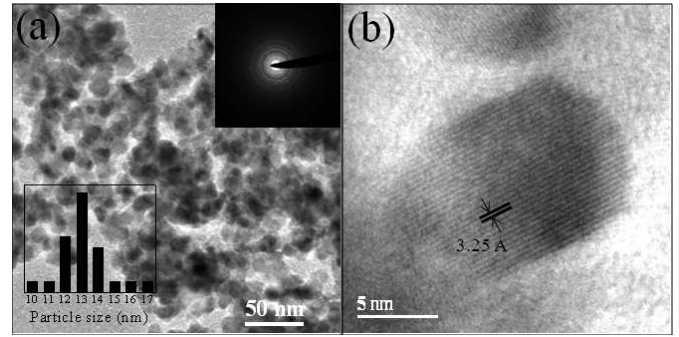}\caption{\label{fig:TEM}(a) TEM image of SnO$_{2}$ nanoparticles. The two
insets show the particle size distribution and the ED pattern, (b)
HRTEM image of a SnO$_{2}$ nanoparticle.}

\end{figure}

Figure \ref{fig:TEM} shows TEM images of SnO$_{2}$ nanoparticles.
Histogram of size distribution is shown in the inset of fig. \ref{fig:TEM}(a).
Average particle size was found to be $\sim13$\,nm which matches
with the value calculated from XRD peak broadening. Rings in electron
diffraction (ED) pattern shown in the inset of fig. \ref{fig:TEM}(a)
confirm the particles to be polycrystalline in nature. Figure \ref{fig:TEM}(b)
shows high resolution TEM image of a SnO$_{2}$ nanoparticle. Lattice
fringes corresponding to the ($110$) planes can be seen clearly.

\begin{figure}
\centering{}\includegraphics[scale=0.3]{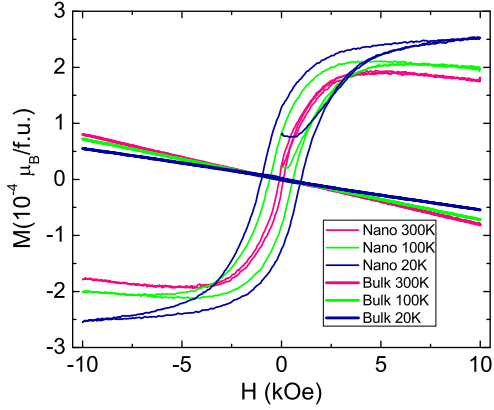}\caption{\label{fig:MHLoop}(Color online) Magnetic hysteresis loops for the
SnO$_{2}$ nano sample (thin curves) and magnetization isotherms for
the bulk sample (thick curves) at various temperatures are shown after
subtracting the diamagnetic holder contribution.}

\end{figure}

Figure \ref{fig:MHLoop} shows magnetic hysteresis loops for the SnO$_{2}$
nano sample as well as magnetization isotherms for bulk SnO$_{2}$
sample at different temperatures. As expected, the bulk SnO$_{2}$
sample simply shows diamagnetic $M$ vs $H$ isotherms. On the other
hand, SnO$_{2}$ nanoparticles show ferromagnetic hysteresis loops
with saturation magnetzation ($M_{S}$) value of $\sim2\times10^{-4}$\,($\mu_{B}$/\textit{formula
unit}). This value was calculated taking the total mass of the nanoparticles.
However, ferromagnetism in SnO$_{2}$ nanoparticles is believed to
be a surface effect and taking only the `surface' mass (corresponding
to a one unit cell thick spherical shell of diameter $13$\,nm),
$M_{S}$ will be $\sim10^{-3}$\,($\mu_{B}$/\textit{formula unit}).

Clearly this bulk technique measures an average of all the magnetic
moments in the sample. In the present case there is bound to be a
distribution of moments; the atoms of the surface having a greater
moment than the core and this distribution possibly being further
dependent on the size of the particles.

\begin{figure}
\centering{}\includegraphics[scale=0.3]{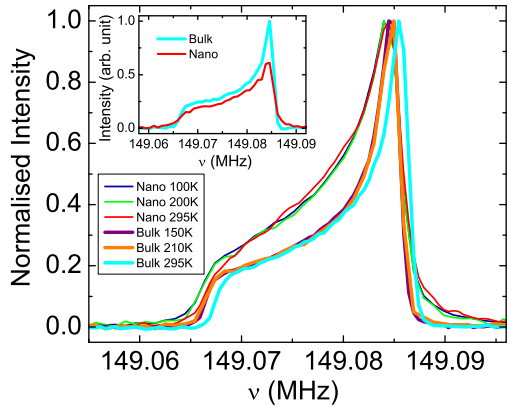}\caption{\label{fig:Spectra}(Color online) Spectra of SnO$_{2}$ bulk (thick
lines) and nano sample (thin lines) at different temperatures. Inset:
Comparison of spectra of bulk and nano sample of SnO$_{2}$ at room
temperature measured with identical parameters and normalised to mass. }

\end{figure}

Our $^{119}$Sn NMR measurements provide a local probe of this magnetism.
The total $^{119}$Sn NMR spectral intensity is found to be smaller
in the nano sample compared to the bulk (see inset of fig. \ref{fig:Spectra})
by about $15\%$. This loss could arise due to static moments on Sn
atoms which provide a large magnetic field at the $^{119}$Sn nucleus
(typical hyperfine coupling $\sim40$$\,$kOe/$\mu_{B}$ \cite{Kojimaa-JMMM-90-1990})
and move the NMR signal out of our window of observation.

The NMR lineshape for the nano sample is somewhat broader than for
the bulk and unchanged with temperature (see fig. \ref{fig:Spectra}).
The increase in width (full width at half maxima increases from about
$4$\,kHz to about $7$\,kHz and total extent of spectra increases
from about $21$\,kHz to about $25$\,kHz) could arise from a distribution
of anisotropy (surface vs core). The total spectral intensity (corrected
for temperature and $T_{2}$ effects) is found independent of temperature
for the nano sample. This implies that the total no. of $^{119}$Sn
nuclei that we probe does not change with temperature or parts of
the sample do not progressively become ferromagnetic as temperature
decreases.

\begin{figure}
\centering{}\includegraphics[scale=0.3]{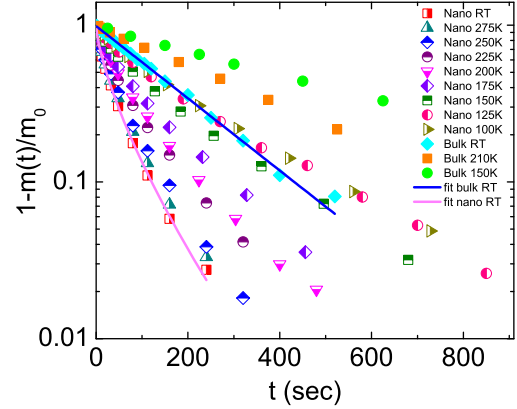}\caption{\label{fig:Recovery}(Color online) Spin-lattice relaxation recovery
curves of the SnO$_{2}$ bulk (solid symbols) and nano sample (half-filled
symbols) at different temperatures. Fitting of the recovery data with
stretched exponential for nano sample and exponential for bulk sample
(both as solid lines) at room temperature (RT) are also shown. }

\end{figure}

\begin{figure}
\centering{}\includegraphics[scale=0.3]{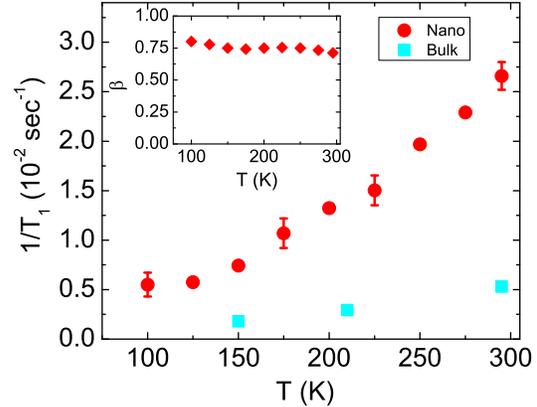}\caption{\label{fig:T1vsT}(Color online) Relaxation rates obtained by fitting
the recoveries of SnO$_{2}$ nano sample with Eq. \ref{eq:T1 equation}
plotted as a function of temperature $T$. The light blue squares
indicate the relaxation rates of the bulk sample. Inset: The temperature
variation of the exponent $\beta$ obtained from fitting of recovery
data with Eq. \ref{eq:T1 equation}.}

\end{figure}

We next focus on the spin-lattice relaxation rate data. The recovery
of the $^{119}$Sn nuclear magnetisation following a $\nicefrac{\pi}{2}$
pulse is shown in fig. \ref{fig:Recovery} for the nano and the bulk
samples at various temperatures. As expected, for the $I=\nicefrac{1}{2}$
$^{119}$Sn nucleus, the nuclear magnetic recovery for the bulk sample
is exponential with an extremely long $T_{1}$ ($\sim180$\,s) at
room temperature. For this sample, the $T_{1}$ increases to about
$560$\,s by $150$\,K and is expected to increase further at lower
temperature. Here, the relaxation probably arises from a coupling
to lattice vibrations typical in nonmagnetic insulators. In contrast,
for the nano sample, the recovery is not single-exponential. However,
we could fit the recovery at all temperatures to a stretched exponential
function \begin{equation}
1-m(t)/m_{0}=exp[-(t/T_{1})^{\beta}]\label{eq:T1 equation}\end{equation}
 where the $\beta$ is nearly unchanged with temperature (see inset
of fig. \ref{fig:T1vsT}). The $1/T_{1}$ thus obtained is plotted
in fig. \ref{fig:T1vsT}. The faster relaxation rate for the nano
sample must then arise from fluctuating paramagnetic moments at Sn
sites. A distribution in the value of such magnetic moments (higher
near the surface, smaller in the core, and perhaps grain size dependent)
then leads to a distribution of $T_{1}$. Note that the longest $T_{1}$
component as obtained from the slope of the data (at $300$\,K for
the nano sample) at high-$t$ in fig. \ref{fig:Recovery} is about
$80$\,s which is still less than half the value for the bulk. For
a paramagnet, the spin-lattice relaxation rate is expected to be nearly
$T$-independent at high-$T$ and might increase a little with a decrease
in temperature. The decrease in $1/T_{1}$ seen by us points to a
gap in the spin excitation spectrum of the local moments. This might
be connected to some novel mechanism (such as orbital magnetism) for
the formation of these moments. There is a possibility that phonon
excitation spectra and the coupling of the nuclei to the phonon bath
get strongly modified in nanoparticles and are still effective in
$1/T_{1}$. However, it seems unlikely that there are only ferromagnetic
and diamagnetic regions in the sample without any paramagnetic ones.
A theoretical effort is needed to understand better the phonons as
also the dynamical susceptibility of defect induced moments in nanoparticles.

\textbf{Conclusions.} - We have probed a nanoparticulate sample of
SnO$_{2}$ using bulk magnetisation and $^{119}$Sn NMR measurements.
While bulk data show hysteresis in the isothermal magnetisation with
magnetic field, evidence of paramagnetic moments is seen in the $^{119}$Sn
NMR $1/T_{1}$ data. The nuclear magnetisation recovery follows a
stretched exponential behaviour with the stretching exponent $\beta\sim0.75$
which is nearly $T$-independent. The relaxation rate is strongly
enhanced in the nano sample in comparison to the bulk. A $15\%$ loss
of the NMR signal in the nano as compared to the bulk is thought to
be from static local moments (due to ferromagnetism) in a part of
the sample which shift the NMR signal out of our window of observation.

\begin{center}
{*}{*}{*}
\par\end{center}

We thank Department of Science and Technology, Govt. of India for
financial support.


\begin{thebibliography}{21}
\bibitem{footnote}\textsc{The longest delay `$t$' used by them in
the $\nicefrac{\pi}{2}-t-(\nicefrac{\pi}{2}-\pi)$ sequence is much
smaller than the $T_{1}$}.
\end{thebibliography}
\end{document}